\documentstyle[preprint,prb,aps]{revtex}  

\begin{document}                
\title{Plasma Waves in Anisotropic Superconducting Films Below and Above the
Plasma Frequency}
\author{ Mauro M. Doria$^a$, Gilberto ~Hollauer$^b$,
F.~Parage$^c$, and O.~Buisson$^c$}
\address{
$^a$ Instituto de F\'{\i}sica, Universidade Federal Fluminense,
C.P. 100.093\\Niter\'oi 24001-970 RJ, Brazil.\\
$^b$ Departamento de F\'{\i}sica, Pontif\'{\i}cia Universidade 
Cat\'olica do Rio de Janeiro, Rio de Janeiro 22452-970 RJ, Brazil.\\
$^c$ Centre de Recherches sur les Tr\`es Basses Temp\'eratures, 
Laboratoire Associ\'e \`a l'Universit\'e\\Joseph Fourier, 
C.N.R.S., BP 166, 38042 Grenoble-C\'edex 9, France.\\ }

\maketitle
\begin{abstract} 

We consider wave propagation inside an anisotropic superconducting film 
sandwiched between two semi-infinite non-conducting bounding 
dieletric media such that along the c-axis, perpendicular to
the surfaces, there is a plasma frequency $\omega_p$ below the superconducting gap.
Propagation is assumed to be parallel to the surfaces in the dielectric medium, 
where amplitudes decay exponentially.
Below $\omega_p$, the amplitude also evanesces  inside
the film, and 
we retrieve the experimentally measured lower 
dispersion relation branch, $\omega \propto \sqrt{\beta}$, and 
the  recently proposed higher frequency branch,
$\omega \propto 1/\sqrt{\beta}$. 
Above $\omega_p$, propagation is of the guided wave type, i.e.,
a  dispersive plane wave  confined 
inside the film that  reflects into 
the dielectric interfaces,
and the modes are approximately described by
$\omega \approx \omega_p \sqrt{ 1
+ (\beta/\beta_0)^2}$, where $\beta_0$ is discussed here.
\end{abstract}

\narrowtext
\section{Introduction}\label{section0}

The experimental measurement of a plasma edge in the infrared reflectivity 
of  High-Tc bulk superconductors\cite{TAMASAKU,GERRITS,KIM} 
$La_{2-x}Sr_xCuO_4$, $YBa_2Cu_3O_{8-x}$\cite{HOMES},
and $Bi_2Sr_2CaCu_2O_8$\cite{TAJIMA,OPHELIA1,OPHELIA2,MATSUDA}
has brought a renewed interest in 
collective oscillations with the Cooper pair density.
Previously, such studies were hampered by the well-known 
argument\cite{ANDERSON} that the Coulomb interaction shifts the 
frequency of such oscilations to above the gap frequency.
Recent theoretical studies\cite{MISHONOV,ARTEMENKO,FERTIG,WEN,BULAEVSKII} 
support the view of  a plasma oscillations along the c-axis, the direction
orthogonal to the  $CuO_2$ planes, caused by a large superconducting gap 
and a high anisotropy in these materials.

In superconducting films, plasma oscillations have distinct properties.
They exist in films regardless of their  anisotropy or layered structure, and regardless of 
the type of pairing or even of the critical temperature value.
This mode was predicted \cite{MOOIJ,MISHONOV2,MIRHASHEN} some time ago
and was bothly observed in   thin  granular aluminium films,
in the hundreds of $MHz$ range\cite{BUISSON},
and in thin  $YBa_2Cu_3O_{8-x}$ films\cite{DUNMORE}, in the
higher frequency range of hundreds of $GHz$.

The study of propagating modes in films is not a new research field, 
they have been  measured since long ago, in several materials ranging 
from metals\cite{BOERSCH,FUKUI,SARID} to 
semiconductors\cite{UL}, and also discussed 
in the theoretical literature\cite{KLIEWER,ECONOMOU,TIEN}.
The novelty for the superconducting film is the very low frequency
where these modes are observed
and the strong temperature dependence, explained by a
collective oscillation with the Cooper pair density.
Recently\cite{DPB} it has been proposed that, similarly to metals
and semiconductors, {\it two } branches of propagating modes  
should  be observed in highly anisotropic superconducting films, e.g.,
the High-Tc compounds.
The experimental observation of these two modes
can provide a method to independently measure  the London
penetration lengths, parallel  and perpendicular  to the film surface.

In this paper we  include into the study 
of mode propagation  in  films 
the existence of a c-axis plasma frequency, 
$\omega_p$, inside the superconducting state.
We only consider the c-axis  perpendicular to the surfaces 
since this configuration is the most easily grown nowadays.
The propagation of plasma modes above and below $\omega_p$, 
their dispersion relations are our goals in the present paper.
We restrict our study to  waves that propagate along the surfaces of
the superconducting film and 
evanesce in the dielectric medium perpendicularly to the 
surfaces. 
Two very distinct behaviors are predicted, namely, 
field amplitudes  that also evanesce inside 
the film away from the interfaces ($\omega < \omega_p$),
and   confined propagation, i.e.,
waves propagate inside the superconductor along an oblique direction 
and  reflect at the dielectric-superconductor interfaces
($\omega > \omega_p$)\cite{YEH}.
In many aspects this last case resembles propagation 
in optic fibers.
In conclusion this paper  considers a plasma frequency $\omega_p$
inside  the superconducting state and studies its effects into 
wave propagation in films.

The study of plasma modes in superconducting films with 
a plasma frequency $\omega_p$ has been previously considered
by Artemenko and  Kobel'kov\cite{ARTEMENKO2} in
the context of kinetic equations for Green functions generalized
to the case of layered superconductors with weak interlayer
coupling.
The present paper provides a more complete study of such modes than theirs
because it shows that there are  symmetric and antisymmetric modes 
for $\omega < \omega_p$ and discusses their several distinct
regimes.
The antisymmetric mode is intimately connected to the transverse current 
component, thus being highly sensitive to the transverse London penetration
length,  in the same fashion that  the symmetric mode  
depends on the longitudinal London penetration length.
We claim  that these  properties can be used to gain
a better understanding of the transverse and longitudinal supercurrent 
components in anisotropic and layered films.
Besides, the present study is done in a  framework distinct from
those authors,  we take the  simplest possible theory,  the London-Maxwell 
theory for anisotropic materials, supposedly valid when all 
lengths are  larger than the inter-layer separation $a$. 
In the context of this theory we  determine  {\it exactly} 
the  mode frequency, $\omega$, as a function of its 
wavelength along the surfaces,  $\beta$, and of the remaining parameters:
the two London penetration lengths,
transverse ($\lambda_{\perp}$) and longitudinal ($\lambda_{\parallel}$)
to the surfaces; the dielectric constant of the non-conducting medium
exterior to the film, $\tilde \varepsilon$; the film thickness $d$; and
the isotropy and frequency independent
$\varepsilon_s$, the simplest phenomenological choice for the 
superconductor's static dieletric constant.

This paper is organized as follows.
In the next section, (\ref{section1}),
we derive the  dispersion relation equations,
above and below the plasma frequency,
using the London-Maxwell theory, and solve them 
{\it exactly}.
Our study is restricted to identical top and bottom dielectric 
media.
Similar  conclusions should also apply to the most general asymmetric case.
Section \ref{section2} deals with the  $\omega < \omega_p$ case,
where we study  the two possible branches, the
so-called symmetric (lower) and antisymmetric (upper) branches, and their
three possible regimes: optical, coupled, and asymptotic.
For each branch we derive, from our exact solution,
useful approximated expressions for each of the above regimes.
In this fashion many of the previous film  studies are shown to be suitable
approximations of our exact solutions.
In section \ref{section3} we study  wave
propagation above $\omega_p$ and, 
obtain, from our exact dispersion relation, the several dispersion 
relation branches found by Artemenko and Kobel'kov\cite{ARTEMENKO2}.
We show  that symmetry still plays an important role and the
two branches below $\omega_p$ split into several ones,
the symmetry of the state being determined by 
the number of half-wavelengths that fit perpendicularly to the film.
In section \ref{section4} we apply the present theory
to the High-Tc superconductors, suitably choosing
the parameter  values.
Finally in section \ref{section5} we summarize our major 
results.
The proof of some of results from our exact solution, such
as the approximated expressions, are the subject of
appendix \ref{apen}.

\section{ The London-Maxwell Theory Applied to Anisotropic 
Superconducting Films} \label{section1}      
In this section we apply the London-Maxwell theory to
describe wave propagation in 
a superconducting film sandwiched between two identical 
non-conducting dielectric media\cite{BUISSON2}. 
An external electromagnetic wave of
frequency $\omega$, and wavenumber $k \equiv \omega/c$,
is absorbed by the superconducting film,
producing a  mode whose dispersion 
relation is $\omega(\beta)$.
We choose a coordinate system where the two plane parallel surfaces 
separating the superconducting film  to the dielectric medium are 
at $x=d/2$ and $x=-d/2$ .
Propagation takes place along the  $z$ axis such that
all fields can be expressed as $F_i(x)\; \exp{[-i(\beta\,z-\omega\,t)]}$. 
At this point we introduce
the time dependence $\exp({i\;\omega\;t})$ to all fields
into the basic equations governing the system.
Current transport inside the film is described by
the first London equation,
\begin{eqnarray}
i\;\omega\;\mu_0 \lambda_{\parallel}^2 \,  
{\bf J}_{\parallel} = {\bf E}_{\parallel}, \qquad
i\;\omega\;\mu_0 \lambda_{\perp}^2 \,
J_{\perp} =  E_{\perp},
\end{eqnarray}
where ${\bf E}_{\parallel}$ and 
${\bf E}_{\perp}$ are the field components
parallel and perpendicular to the film surfaces, 
respectively.
The electromagnetic coupling, given by
Maxwell's equations,
\begin{eqnarray}
{\bf \nabla} \cdot \;{\bf D} &=& 0 \\ 
{\bf \nabla} \cdot\; {\bf H} &=& 0\\ 
{\bf \nabla} \times \;{\bf E} &=& - i\omega \mu_0 {\bf H}\\
{\bf \nabla} \times \; {\bf H} &=&  i \omega {\bf D} \quad \mbox{where} \quad
{\bf D} = \epsilon_{s} {\bf E} -i{\bf J}/\omega. 
\end{eqnarray}    
shows that the  superconductor  dielectric constant is
tensorial, ${\bf D} = \epsilon_0 {\bf \varepsilon} \cdot {\bf E} $.
\begin{eqnarray} 
{\bf \varepsilon}  =  
\pmatrix{ \varepsilon_{\perp}  & 0 & 0  \cr 0 & \varepsilon_{\parallel}& 0  
\cr 0 & 0 & \varepsilon_{\parallel}  \cr } \quad
\varepsilon_{\perp}= \varepsilon_{s}- {1 \over {(k\lambda_{\perp})^2}} \quad
\varepsilon_{\parallel}=\varepsilon_{s}-{1\over {(k\lambda_{\parallel})^2}}, 
\end{eqnarray}

Phenomenological  theories, such as the present, describe the superconductor 
only at energies much lower than the pair breaking threshold.
The frequency $\omega$ is much smaller
than the frequency defined by the superconducting gap.
Consequently the frequency is much smaller than the plasma frequency along the 
$CuO_2$ planes,
$ k\lambda_{\parallel} \ll 1$, thus rendering a negative, 
and large in modulus, dielectric tensor component parallel to the surfaces.
\begin{eqnarray}
|\varepsilon_{\parallel} | \gg 1 \quad
\varepsilon_{\parallel} \approx -{{1}\over{(k\lambda_{\parallel})^2}}  
\label{cond}
\end{eqnarray}
In the present model the c-axis plasma frequency is reached when
the dielectric tensor component perpendicular to the surfaces becomes null.
\begin{eqnarray}
\varepsilon_{\perp}(\omega=\omega_p)=0 \qquad \omega_p = 
{{c}\over {\sqrt{\varepsilon_s}\lambda_{\perp}}}  \label{plo}
\end{eqnarray}
Hence $\varepsilon_{\perp}$ changes sign  
as $\omega$ crosses $\omega_p$.

Solving Maxwell's equations for this particular geometry
gives two independent sets of field components,
$(H_x,E_y,H_z)$, the transverse electric (TE), and  $(E_x,H_y,E_z)$, the 
transverse magnetic (TM) propagating modes.
For  the TM mode the interfaces superconductor-dielectric acquire 
superficial charge densities. 
Curiously anisotropy also implies in a volumetric charge  density 
inside the film, 
$ \rho = -i {\bf \nabla} \cdot {\bf J}/ \omega = 
\epsilon_0 \big(1-  \varepsilon_{\perp}/ \varepsilon_{\parallel} \big )
\partial E_{\perp}/\partial x$, although
it is not responsible
for these low frequency propagating modes.
In fact such modes
were first observed in nearly isotropic thin granular aluminum films
\cite{BUISSON}.
In conclusion the TM mode supports low frequency travelling waves because
of the  oscillating superficial charge densities
that couple the non-conducting dielectric medium to
the superconducting film,
their ratio given by,
\begin{eqnarray}
{{\sigma (x=d/2)}\over {\sigma (x=-d/2)}} =
- {{\partial E_z/\partial x |_{x=d/2}}\over
{\partial E_z/\partial x |_{x=-d/2}}}
\end{eqnarray}
The TM field equations are given below,
for the dielectric medium, ($x \ge d/2$ and $x\le -d/2$),
\begin{eqnarray}
E_x = i{{\beta} \over {\tilde \tau^2 }} 
\;{{\partial E_z}\over{\partial x}}, \quad 
H_y = i\,\epsilon_0\,{{\omega\, \tilde \varepsilon}\over{\tilde \tau^2}} \;
{{\partial E_z}\over{\partial x}}, \quad
{{\partial^2 E_z}\over{\partial x^2}} - {\tilde \tau}^2\, E_z = 0 , \quad {\tilde \tau}^2 = \beta^2 - k^2  \tilde \varepsilon
\end{eqnarray}
and for the superconducting strip ($ -d/2 \le x \le d/2$),
\begin{eqnarray}
E_x = i{{\beta \, \varepsilon_{\parallel}  } \over 
{\tau^2 \, \varepsilon_{\perp} }} \;{{\partial E_z}\over{\partial x}}, \quad 
H_y = i\, \epsilon_0\,{{\omega\, \varepsilon_{\parallel}}\over{\tau^2}} \;
{{\partial E_z}\over{\partial x}}, \quad
{{\partial^2 E_z}\over{\partial x^2}} - \tau^2\, E_z = 0
\quad \tau^2 = {{ \varepsilon_{\parallel}}\over{\varepsilon_{\perp}}}\beta^2 
- k^2  \varepsilon_{\parallel}  
\end{eqnarray}

In this paper we seek  propagating modes that
evanesce in the dielectric medium, namely
display  exponential decay in the dielectric.
This  condition is ${\tilde \tau}^2 = \beta^2[1 - (\omega/\beta v)^2] >0$,
and can be interpreted as demanding a phase velocity,
$\omega/\beta$,   smaller than the speed of light in the dielectric, 
$ v \equiv c/\sqrt{\tilde \varepsilon} $. 
Hence above  the film ($x \ge d/2 $) one gets that,
\begin{eqnarray}
E_z = \tilde E_o \exp{(-\tilde \tau\;x)} \label{ex}
\end{eqnarray}
From the other side exponential decay inside the film is not sure
to occur because the sign of $\tau^2$ is not uniquely defined. 
This sign determines different physical regimes, and for
this reason, we introduce the below and above plasma
condition and study their respective $\tau^2$ regions.
Hereafter, in order to simplify further discussions, 
we will choose the nonconducting dielectric medium to have
the largest static constant, $\tilde \varepsilon > \varepsilon_s$.
\vskip 0.5truecm
\noindent \par $\underline{\omega<\omega_p}$ \quad In this case both
$\varepsilon_{\parallel}$ and $\varepsilon_{\perp}$ have the same sign. 
Introducing the Eq.(\ref{cond}) approximation, one gets that
$\tau^2 \approx |\varepsilon_{\parallel}/\varepsilon_{\perp}|\beta^2 + 
1/\lambda_{\parallel}^2>0$.
The planes of constant phase are $z=const.$ and field amplitudes also 
evanesce  inside the superconductor.
\begin{eqnarray}
E_z = E_o \exp{(-\tau\;x)} + F_o \exp{(\tau\;x)}, \quad \tau^2>0
\label{deftau}
\end{eqnarray}
\vskip 0.5truecm  
\noindent \par $\underline{\omega >\omega_p}$ \quad In this case 
$\varepsilon_{\parallel}<0$ and $\varepsilon_{\perp}>0$.
Even within  the Eq.(\ref{cond}) approximation,
$\tau^2 \approx -|\varepsilon_{\parallel}/\varepsilon_{\perp}|\beta^2 
+ 1/\lambda_{\parallel}^2$ has no definite sign.
The mode may be evanescent or propagative inside the superconductor.
However the evanescent modes disappear for 
$\tilde \varepsilon > \varepsilon_s$ and  then one gets that $\tau^2<0$.
In this case  the planes of constant phase are 
$\beta\, z \pm \tau'\, x = const.$ ($\tau'^2= -\tau^2>0$), and 
we fall into the case of confined propagation, with an oblique incidence 
well defined at the interfaces.
\begin{eqnarray}
E_z = E_o \exp{(-i\tau'\;x)} + F_o \exp{(i\tau'\;x)}, 
\quad \tau'^2 = -{{\varepsilon_{\parallel}}\over{\varepsilon_{\perp}}}\beta^2 
+ k^2  \varepsilon_{\parallel}>0  \label{deftau'}
\end{eqnarray}
In summary for the present purposes the sign of $\tau^2$ is uniquely 
determined below and above $\omega_p$
by the sign of the ratio $\varepsilon_{\parallel}/\varepsilon_{\perp}$.

The dispersion relations follow from the continuity of the 
ratio  $H_y/E_z$ at a single interface, say $x=d/2$, once 
assumed  that the longitudinal field $E_z$  
is either an even or an odd function with respect to the $x=0$ plane. 
This is possible because the two dielectric media, above
and below the film, are equal.
\begin{center}
Table I~: The four possible dispersion relations and $E_z$.\\
\begin{tabular}{|c|c|}
 $E_z$ & {\it Dispersion Relation}  \\ \hline 
 $E_{oz}\; \cosh{(\tau \;x)}$ &  
${{\tau\;\tilde\varepsilon}\over{\tilde\tau\;\varepsilon_{\parallel}}} 
= -\tanh{(\tau\;{d\over 2})}$  \\
$E_{oz}\; \sinh{(\tau \;x)}$ &  
${{\tau\;\tilde\varepsilon}\over{\tilde\tau\;\varepsilon_{\parallel}}} 
= -{1\over{\tanh{(\tau\;{d\over 2})}}}$  \\
$E_{oz}\; \cos{(\tau' \;x)}$ &  
${{\tau'\;\tilde\varepsilon}\over{\tilde\tau\;\varepsilon_{\parallel}}} 
= -\tan{(\tau'\;{d\over 2})}$  \\
$E_{oz}\; \sin{(\tau' \;x)}$ &  
${{\tau'\;\tilde\varepsilon}\over{\tilde\tau\;\varepsilon_{\parallel}}} 
= {1\over{\tan{(\tau'\;{d\over 2})}}}$  
\end{tabular}
\end{center}
\begin{eqnarray}
\label{tab1}
\end{eqnarray}
At this point we  introduce some  dimensionless variables
useful in the study of the above dispersion relations.
\begin{center}
Table II~: Dimensionless Variables.\\
\vspace{0.5cm}
\begin{tabular}{|c|c|c|c|c|}
 Retardation  & Anisotropy  & Dielectric 
 & Thickness & Wave Number \\ \hline 
$\gamma \equiv {{\omega/\beta}\over{v}}$  &  
$ R \equiv \big({{\lambda_{\perp}}\over{\lambda_{\parallel}}} \big)^2$ &
$ r \equiv {{\tilde \varepsilon}\over{\varepsilon_s}}$ &
$ A \equiv \big({{\lambda_{\parallel}}\over{d}}\big)^2$ &

$ X = \big ( \beta \; \lambda_{\parallel} \big)^2  $
\end{tabular}
\end{center}
\begin{eqnarray}
\label{tab2}
\end{eqnarray}
We cast some of our previous results in the dimensionless variables
formalism.
The condition of  evanescence along the direction orthogonal to the surfaces 
is $0 \le \gamma \le 1$, since according to Eq.(\ref{ex}), one has that
$E_z = \tilde E_o \exp{(-\sqrt{X \over A}\;\sqrt{1-\gamma^2}x/d)}$.
The dominance of   the incident wavelength over London's penetration along 
the surface, which gave Eq.(\ref{cond}), becomes $\gamma^2 \ll r/ X $.
The dielectric component ratio,
within the  Eq.(\ref{cond}) approximation, is
$\varepsilon_{\perp}/\varepsilon_{\parallel} \approx (1/ R)\, 
\big( 1 - R\, \gamma^2 \, X/r \big)$.
Thus the $\omega<\omega_p$ and  $\omega>\omega_p$ regimes correspond to
$R\, \gamma^2 \, X /r < 1 $ and  $R\, \gamma^2 \, X /r > 1 $, respectively.
The c-axis (High-Tc compounds)  is  perpendicular to
the surfaces and so $R  \ge 1$.

In the following, we  obtain the {\it exact} solution for each of the Eq.(\ref{tab1}) 
dispersion relations considering
the dimensionless variables $d\,\tau$ ($\omega<\omega_p$) and 
$d\,\tau'$ ($\omega>\omega_p$) as a parameter $s$, whose
range is still to be
determined below and above $\omega_p$.
Therefore we  seek the dimensionless curve, $\gamma(X)$,
in parametric form, $[X(s),\gamma(s)]$, and
immediately obtain the  dispersion relation,
$[\beta(s),\omega(s)] =[\sqrt{X(s)}/\lambda_{\parallel}, 
(v/\lambda_{\parallel})\,\gamma(s) \,\sqrt{X(s)}]$.

\vskip0.5truecm
\par \noindent $\underline{\omega < \omega_p}$ \quad We introduce the dimensionless parameter,
\begin{eqnarray}
t \equiv d\,\tau = {1 \over{\sqrt{A}}}
\sqrt{ {{X\,R}\over{1-{{\gamma^2\,X\,R}\over{r}}} }+1}  \label{eqt}
\end{eqnarray}

Using the Eq.(\ref{tab2}) dimensionless variables, the  Eq.(\ref{tab1}) 
dispersion relations become, 
\begin{center}
Table III~: The  dispersion relations for $\omega < \omega_p$ \\
\begin{tabular}{|c|c|}
 $E_z$ & {\it Dispersion Relation}  \\ \hline 
 $E_{oz}\; \cosh{(t x / d)}$ &  
$\sqrt{A\;X} {{\gamma^2}\over{\sqrt{1-\gamma^2}}} 
= {{\tanh{(t/2)}}\over t}$  \\
$E_{oz}\; \sinh{(t x / d)}$ &  
$\sqrt{A\;X} {{\gamma^2}\over{\sqrt{1-\gamma^2}}} 
= {1\over {t\;\tanh{(t/2)}}}$  
\end{tabular}
\end{center}
\begin{eqnarray}
\label{tab3}
\end{eqnarray}
To find their solution, first obtain
$X(t,\gamma)$ from Eq.(\ref{eqt}),
and then  introduce it into
Eq.(\ref{tab3}). 
We obtain a second degree equation
for $\gamma(t)^2$ with two roots: one negative the other positive.
The negative root does not correspond to propagating modes and only the positive is left.
\begin{eqnarray}
\gamma_{I}(t)^2 &=& {{(A\,t^2-1)-r+\sqrt{[(A\,t^2-1)+r]^2\;+\; 
4\;{{A\,r^2}\over{R}}\;(A\,t^2-1)\;g_{I}(t)} }\over
{2(A\,t^2-1)[1\;+\;{{A\,r}\over{R}}\;g_{I}(t)]}}   \label{gbpl} \\
X(t) &=& {{r}\over{R}}{{A\,t^2-1}\over{\gamma_{I}(t)^2\;(A\,t^2-1)+r}} 
 \label{xbpl}  
\end{eqnarray}   
$I=\{S,A\}$ labels the $E_z$ symmetry: $g_{A}(t) = t^2\,\tanh^2{(t/2)}$ and 
$g_{S}(t)= t^2/\tanh^2{(t/2)}$  are associated to
the antisymmetric and the symmetric  modes, respectively.
According to the above equations the parameter $t$ range is 
[$1/\sqrt{A},\infty$].
\vskip0.5truecm
\par \noindent $\underline{\omega > \omega_p}$ \quad
Through Eq.(\ref{deftau'}) we define 
a new dimensionless parameter.
\begin{eqnarray}
t' \equiv d\,\tau' = {1 \over{\sqrt{A}}}
\sqrt{ {{X\,R}\over{ {{\gamma^2\,X\,R}\over{r}}-1 } }-1}  \label{eqt'}
\end{eqnarray}
Introducing the  Eq.(\ref{tab2}) dimensionless variables into
the Eq.(\ref{tab1}) dispersion relations gives that,
\begin{center}
Table IV~: The  dispersion relations $\omega>\omega_p$ \\
\vspace{0.5cm}
\begin{tabular}{|c|c|}
 $E_z$ & {\it Dispersion Relation}  \\ \hline 
 $E_{oz}\; \cos{(t' x / d)}$ &  
$\sqrt{A\;X} {{\gamma^2}\over{\sqrt{1-\gamma^2}}} 
= {{\tan{(t'/2)}}\over t'}$  \\
$E_{oz}\; \sin{(t' x / d)}$ &  
$\sqrt{A\;X} {{\gamma^2}\over{\sqrt{1-\gamma^2}}} 
= -{1\over {t'\;\tan{(t'/2)}}}$  
\end{tabular}
\end{center}
\begin{eqnarray}
\label{tab4}
\end{eqnarray}
To solve them, express Eq.(\ref{eqt'}) as  $X(t',\gamma)$ and introduce  
it back into Eq.(\ref{tab4}) obtaining
a second degree equation for $\gamma(t')^2$,
whose solution is,
\begin{eqnarray}
\gamma_{I}(t')^2 &=& {{(A\,t'^2+1)+r \pm \sqrt{[(A\,t'^2+1)-r]^2\;-\; 
4\;{{A\,r^2}\over{R}}\;(A\,t'^2+1)\;h_{I}(t')} }\over
{2(A\,t'^2+1)[1\;+\;{{A\,r}\over{R}}\;h_{I}(t')]}}   \label{gapl} \\
X_{I}(t') &=& {{r}\over{R}}{{A\,t'^2+1}\over{\gamma_{I}(t')^2\;(A\,t'^2+1)-r}} 
\label{xapl}
\end{eqnarray} 
Like in the previous case, $I=\{S,A\}$  gives 
the $E_z$ parity:
$h_{A}(t') =t'^2\,\tan^2{(t'/2)}$  and
$h_{S}(t') = t'^2/\tan^2{(t'/2)}$ for the
 antisymmetric and symmetric modes, respectively.

The discussion of the suitable parameter range
is more involving in the present case.
According to Eq.(\ref{tab4}) the function 
$t'\tan{(t'/2)}$ must remain negative for the antisymmetric
problem whereas $\tan{(t'/2)}/t'$ 
must be positive for the symmetric case.
This restricts  the parameter range $t' $ to  the intervals, 
$[2 \, N \, \pi, (2 \, N+1) \, \pi]$ for symmetric and 
$[(2 \, N+1) \, \pi, 2\, (N+1) \, \pi]$ for antisymmetric,
where $N$ is an integer  larger or equal to zero.

Our  {\it exact} parametrized solutions $\omega(\beta)$ are
ready for applications, just requiring the numerical parameters.
Before doing so  in section \ref{section4}, 
we find convenient to  describe their physical properties
in the next two sections.

\section{ Dispersion Relation Below the Plasma Frequency}\label{section2}     

In this section we study the properties of propagating modes below $\omega_p$.
We find convenient to  summarize their major physical properties and
introduce  some approximated expressions, each describing a different regime
of the exact  $\omega < \omega_p$ dispersion curve. 
We leave to the appendix \ref{apen} the proof that
the previous section exact results do
justify our picture and yield the approximated expressions.
Below the plasma frequency  the field amplitudes inside the film
evanesce from the surfaces, and this exponential decay is characterized 
by $\tau$,  according to Eq.(\ref{deftau}).
The choice of film thickness, $d$, is very
important in order to assure sufficient coupling between
the two surfaces. 
For extremely thick films the surfaces decouple and the
symmetric and the antisymmetric modes are the same, 
only independent surface plasma modes exist in this situation.
Indeed, in  case of strong coupling between the two surfaces 
three distinct regimes are possible for both symmetric and antisymmetric 
modes.
In sequence of increasing $\beta$ we call them {\it optical}, {\it coupled} 
and {\it asymptotic} (see Fig.(\ref{fig2})).

Close to  the origin $\beta\approx 0$, the  modes are optical, that is, they
are essentially plane waves travelling in a dielectric
medium. 
There is almost no exponential decay in the surrounding dielectric 
semi-spaces up to appreciable distances,
$\tilde \tau \approx 0$, according to Eq.(\ref{ex}).
Propagation occurs with the speed of light in the dielectric,
($\gamma \approx 1$). 
The kinetic energy of the superconducting carriers is negligible compared 
to the magnetic energy of the mode. 
The superconducting film contributes very weakly to this regime.
In the optical regime one has, approximately, the linear 
behavior $\omega \approx v\,\beta$.

Slowly the condensate's kinetic energy increases
with $\beta$ until finally both energies
become comparable and a first cross-over takes place.
This is the onset of the  {\it coupled}  regime, a slow mode ($\gamma <1$),
where the condensate's kinetic energy dominates over the magnetic energy,
the film and the dielectric are strongly coupled. 
This is the true plasma mode regime.
Such evanescent  propagating modes are known
\cite{MOOIJ,MIRHASHEN,MISHONOV2,DPB},
here we discuss  the effects of a c-axis  plasma frequency 
$\omega_p$ below the gap\cite{ARTEMENKO2}.
In the coupled regime inside the film, the fields evanesce from the 
surfaces very smoothly resulting into
two types of coupling between the superficial plasmons,
which are the antisymmetric and  the symmetric modes.
For the symmetric mode  its dispersion relation has been 
theoretically studied by many authors\cite{MOOIJ,MIRHASHEN,MISHONOV2} 
and experimentally 
observed\cite{BUISSON,DUNMORE} in the past:
\begin{eqnarray}
\omega_s \;\approx\;  {{v}\over{\lambda_{\parallel}}}\; 
\sqrt{{{d\,\beta}\over 2}}    \label{eqs}
\end{eqnarray}
Recently\cite{DPB} it has been proposed that 
the symmetric mode is just the lowest frequency
branch. In fact, there is an upper (antisymmetric) branch mode,
that  can also be experimentally observed 
for highly anisotropic High-Tc 
superconducting films:
\begin{eqnarray}
\omega_a \;\approx \;{{v}\over{\lambda_{\perp}}}\; 
\sqrt{{{1}\over{d\,\beta/2+\varepsilon_s/\tilde\varepsilon}}}\label{eqa}
\end{eqnarray}
Because the antisymmetric branch is the highest in frequency, it is more
sensitive to the effects of the  plasma frequency $\omega_p$.
For $d\beta/2$ comparable (or smaller) to $\varepsilon_s/\tilde \varepsilon $, 
this branch becomes difficult to observe since
$\omega_a \approx \omega_p$.
For $d\beta/2$ dominant over $\varepsilon_s/\tilde \varepsilon $, 
the symmetric and antisymmetric relations are proportional to
$\sqrt{\beta}$ and $1/\sqrt{\beta}$, respectively.

The cross-over frequency between the optical and the symmetric regime is 
obtained at the intersection of the two branches:
\begin{eqnarray}
\omega_{cross,s}=  {{v d}\over{2\lambda_{\parallel}^2}} \label{crs} 
\end{eqnarray}
The cross-over between optical and antisymmetric modes is implicitly given by
the unique real solution:
\begin{eqnarray}
({d \over{2 v}}) \,\omega_{cross,a}^3+
({\varepsilon_s \over {\tilde\varepsilon}})\, \omega_{cross,a}^2-
({{v} \over {\lambda_{\perp}}})^2=0 \label{cra}
\end{eqnarray}
For the antisymmetric mode, the optical and the coupled  frequencies are, 
respectively,  an increasing and a decreasing function of $\beta$. 
Thus the cross-over $\omega_{cross,a}$ also gives an estimate of the maximum 
frequency, associated to the peak seen in the antisymmetric curve 
(Fig.(\ref{fig2})).

By increasing $\beta$, evanescence inside
the superconductor becomes stronger ($\tau$ is large),
up to the point where the surfaces are nearly decoupled.
We have just reached the asymptotic regime.
Thus for sufficiently high $\beta$, symmetric and antisymmetric modes  
converge to the same asymptotic frequency, 
\begin{eqnarray}
\omega_{as} =    \omega_p \;
\sqrt{\big({{1+\sqrt{1+4\,   
\big({{\tilde\varepsilon\, \lambda_{\parallel}}\over
{\varepsilon_s \,\lambda_{\perp}}}\big)^2}}
\over{1+2 \big({{\tilde\varepsilon\, \lambda_{\parallel}}\over
{\varepsilon_s \,\lambda_{\perp}}}\big)^2+
\sqrt{1+4 \big({{\tilde\varepsilon\, \lambda_{\parallel}}\over
{\varepsilon_s \,\lambda_{\perp}}}\big)^2}}} \big)} \label{omas}
\end{eqnarray}
Notice that it is always true that $ \omega_{as} <\omega_p$. 
It is easy to check in  the following extreme cases of
a very high  dieletric constant 
($ \tilde \varepsilon/ \varepsilon_s \gg 
\lambda_{\perp}/\lambda_{\parallel}$ ), and of the opposite case, namely,
when anisotropy plays a more important role than 
the dielectric constant, 
($ \tilde \varepsilon/ \varepsilon_s \ll \lambda_{\perp}/\lambda_{\parallel}$ ).
In the former case we obtain\cite{DPB}
$\omega_{as} = v/\sqrt{\lambda_{\perp}\lambda_{\parallel}}$, and in 
the latter  $\omega_{as}=\omega_p$.
In summary we have just reviewed the major features of 
the three possible regimes for both the
symmetric and the antisymmetric dispersion relations.

We find that the dielectric constant
ratio  $r$ must be sufficiently high in order to prevent
that the $\beta$ range be extremely limited for the {\it coupled} regime.
In the next section (\ref{section4}) we study such crossovers in case of
the High-Tc parameter values.
The criterion for the disapperance of the
coupled antisymmetric and symmetric regimes is just given by
an asymptotic frequency $\omega_{as}$  smaller than $\omega_{cross,s}$
and $\omega_{cross,a}$, respectively.
As an example consider the case of an extremely large anisotropy 
($\lambda_{\perp} \rightarrow \infty$), where we get that
$\omega_{as} \approx \omega_p \rightarrow 0$, eventually smaller
than the crossover frequencies of Eq.(\ref{crs}) and Eq.(\ref{cra}).
In this limit the dispersion relation goes almost directly from 
the optical to the asymptotic regime.


\section{ Dispersion Relation Above the Plasma Frequency}\label{section3}

In this section propagating modes inside
the superconducting film with the frequency larger
than $\omega_p$ are considered. 
We summarize their major physical properties, 
and leave to the appendix \ref{apen} their  derivation from 
the exact parametrized dispersion relation. 
We are again interested in modes that evanesce in the dieletric medium.
As discussed before, $\tau'$ is always real with the experimental 
parameter choice of $\tilde \varepsilon > \varepsilon_s$.
The propagation of light inside the film can be understood
by a superposition of two dispersive plane waves,
$\exp{ \{-i[\pm \tau'(\beta) x \,+\, \beta z \,-\, \omega(\beta) t]\} }$,
regarded as incident and reflected waves, 
thus with a well defined angle of incidence at the interfaces 
superconductor-dielectric.
Thus while outside the film light is still evanescent,
in its interior the propagation can be pictured
through ray optics (see Fig.(\ref{fig1})), similarly to an optical fiber.

The exact parametrized plasma relation, given  by  Eq.(\ref{gapl}) and 
Eq.(\ref{xapl}), leads to the
simple approximated dispersion relation, 
in the vicinity of $\omega/\beta \approx v$,
\begin{eqnarray}
\omega_M = \omega_p \sqrt{1 + \;
{{(\beta\,d)^2}\over{(\pi M)^2 + (d/\lambda_{\parallel})^2}}\,
\big({{\lambda_{\perp}}\over{\lambda_{\parallel}}}\big)^2
  } \label{obap}
 \end{eqnarray}
where M is a positive integer, odd for the symmetric modes ($M=2N+1$) and 
even for the antisymmetric modes ($M=2(N+1)$).
This approximated expression was first obtained by Artemenko and Kobel'kov
\cite{ARTEMENKO2}.
The $M^{th}$ curve satisfies the evanescence condition, 
$\gamma \le 1$, for  $\beta > \beta_{init}(M)$.
\begin{eqnarray}
{\beta_{init}(M)}=\ {{1}\over{\lambda_{\perp}\sqrt{{{\varepsilon_s}\over{\tilde \varepsilon}}- {{1}\over{(\pi M)^2}}
\big({{d}\over{\lambda_{\parallel}}}\big)^2}}}
\end{eqnarray}
At the particular value $\beta =\beta_{init}(M)$, the wave vector 
along the x-direction
is given by $\tau'\, d = M \, \pi$ where $M$ is exactly  the number of 
half-wavelength that fit perpendicularly to the film. 
For $\beta >\beta_{init}(M)$, the relation between 
$\tau'$ and $M$ is not so simple because the wave is also in the dielectric
medium. In this case $M$ just determines the number of $E_z$ extrema along
the x-direction (see Fig.(\ref{fig1})).
For instance, when just one half-wavelength fits into
the film we are facing the
$M=1$  symmetric mode, which has just a single $E_z$ extremum (maximum).
We notice that the frequency of the plasma modes decreases for increasing M. 
For $\beta <\beta_{init}(M)$  we have that
$\gamma > 1$ and the modes are not evanescent,
they are plane waves travelling in the dielectric medium. 
They have an oblique incidence at the surfaces and 
to see this just take the plane wave  
$\exp{ \{-i[\tilde \tau x + \beta z - \omega t]\} }$ with 
the speed of light in the dielectric,
$v = \omega/\sqrt{\tilde\tau^2 +\beta^2}$.
One obtains  that $\tilde\tau = \beta \sqrt{\gamma^2-1}$, showing that we 
are in a $\gamma$ regime not studied here.

The present theory eventually breaks down because large $\tau'$ means 
an infinitesimally small wavelength inside the film along the $x$ axis.
For this reason  a physical cut-off must be put into this theory, the 
inter-plane separation $a$. 
The upper limit $\tau_{max}'$ is determined when a x-direction 
half-wavelength fits into the interplane  distance $a$.
\begin{eqnarray} 
\tau_{max}'=\pi/a=M_{max}\pi /d \label{Max}
\end{eqnarray}
Introducing $M_{max}$ in the approximated dispersion relation 
Eq.(\ref{obap}), gives the lowest frequency plasma mode:
$\omega_{min} = \omega_p\, \sqrt{1+(\beta\lambda_J/\pi)^2}$, 
where $\lambda_J \equiv a \lambda_{\perp}/\lambda_{\parallel}$
is the  Josephson penetration depth.

Beyond $\tau_{max}'$  more elaborate models should
provide a description of the condensate in such scale.
This can render the theory's usage quite limited, in case the ratio
$d/a$ is small, because for very few half-wavelengths inside the film
we reach the cut-off limit.
However the applicability of this model above $\omega_p$ is not restricted
to a small ratio $d/a$ because
the propagating mode has constant amplitude inside
the  film, and so, no matter how far apart
the surfaces are, there will always be  waves reflected at the surfaces.
For this reason  the major conclusions of this section, 
namely, the existence of slow modes above the plasma frequency,
corresponding to dispersive plane waves inside the film,
must remain valid even near the cut-off limit.

In the next section we take the standard 
parameter values for the high-Tc material and discuss 
the properties of such propagating modes
below and above the plasma frequency.
 

\section{ Applications}\label{section4}

For our applications we choose the following set of parameters 
for the High-Tc ceramic superconductors.
The static dielectric constant \cite{TAMASAKU} is $\varepsilon_{s}\approx 30$.
The zero-temperature  London penetration length along the $CuO_2$ planes
is  $\lambda_{CO}= 0.15 \mu m$ \cite{BLATTER}.
While  the anisotropy,
$\lambda_{\perp}/\lambda_{\parallel}$, is  $5$ for 
$YBa_2Cu_3O_{8-x}$\cite{BLATTER}, 
it has been changing  in
the past for $Bi_2Sr_2CaCu_2O_x$, ranging from $10^2$, mainly from
torque measurements studies \cite{BLATTER}, to much higher values\cite{BRANDT}.
We show here that choosing the anisotropy between these two
values can have important effects on the properties of
the dispersion relations for $\omega < \omega_p$.
There is  also the compound $Tl_2Ba_2CaCuO_x$ compound\cite{GRAY} with 
$\lambda_{\perp}/\lambda_{\parallel}\approx 90$.

Fig.(\ref{fig1})  provides  
a pictorial intuitive view of wave propagation in the film below and
above the plasma frequency. 
Below $\omega_p$ the instantaneous
electric field, as well as the component  $E_z$,
are shown for both symmetric and antisymmetric fields.
The surface charges  are the sole sources of  propagating electric 
fields.
The superficial charge arrangement
has strong consequences for the electric
field distribuiton inside the film, leading to symmetric and antisymmetric 
modes, found at distinct frequency ranges, the latter being an upper
mode.
To understand the effects of anisotropy, or a layered
structure, into these modes,
where the  $CuO_2$  planes are parallel to the film surface, 
consider  the  transverse and longitudinal field  and  
supercurrent components. 
$E_x$ and $J_x$ are very  intense for the antisymmetric
mode and nearly zero for the symmetric one, and
$E_z$ and $J_z$,  are the dominant components for the symmetric mode 
but not for the antisymmetric one.
Thus the relevant penetration depths for the
symmetric and the antisymmetric modes must be
$\lambda_{\parallel}$ and $\lambda_{\perp}$, respectively.
The anisotropy  $\lambda_{\perp}>\lambda_{\parallel} $ 
hardens the system  along the  c-axis  thus making the
antisymmetric mode lower in frequency.
Above $\omega_p$ a dispersive
plane waves travels inside the film that undergoes total reflection
at the interfaces.
This Figure also depicts the number of extrema for the electric component  
$E_z$, which 
determines the symmetry of the  mode.

In Fig.(\ref{fig2}), the symmetric and antisymmetric dispersion relations are
shown for a very thin film $d=10nm$ and the three anisotropies,
$\lambda_{\perp}/\lambda_{\parallel}=5$, $10^2$ and $10^3$.
In order to slow down as much as possible
light in the dielectric, and so,  lower the coupled regime frequency range,
we choose $SrTiO_3$ as the exterior  non-conducting dielectric
medium. 
At low temperatures its dielectric constant is known to be high up to the
GHz frequency \cite{BUISSON}, $\tilde \varepsilon \approx 20000$.
All curves in this Figure were directly obtained from the parametrized
solution of Eq.(\ref{gbpl}) and Eq.(\ref{xbpl}).
The symmetric state crossover frequency, $\omega_{cross,s}$, 
between the optical and the square root regimes is found to be $75\;GHz$ 
for the three anisotropies displayed here.
The antisymmetric state crossover frequency, $\omega_{cross,a}$, 
between the optical and the coupled regime, depends on the anisotropy, 
being $2.3\;THz$, $290\;GHz$ and $48\;GHz$ for 
$\lambda_{\perp}/\lambda_{\parallel}=5$, $10^2$ and $10^3$, respectively.
Because of this anisotropy dependence they are not indicated in this Figure.
Notice that the symmetric mode (dashed lines) is always found at 
a frequency range lower than its corresponding
antisymmetric mode (continuous lines).
In the low wave vector limit, all the dispersions collapse into the same 
linear (optical) regime.
In the opposite limit, $\beta$ very large, both symmetric and antisymmetric
curves converge to the same asymptotic frequency.
This asymptotic frequency is just the surface plasma frequency, 
strongly affected by  anisotropy: $1\; THz$ for $YBCO$, $200\; GHz$ for $BSSCO$ 
and $50 \; GHz$ for $\lambda_{\perp}/\lambda_{\parallel}$=$10^3$. 
This saturation frequency is independent on film thickness, as expected,
according to  Eq.(\ref{omas}).
Notice that the wave vector signaling the onset of the asymptotic regime 
decreases for increasing anisotropy, ranging from $10 \mu m^{-1}$ 
for YBCO to  $100 mm^{-1}$ for the maximum anisotropy considered here.
Between the optical and the asymptotic is the coupled regime,
the true plasma mode,
found to  extend over
a large range of frequency and wavenumber for an  $YBCO$ thin film,
according to this Figure.
In this regime the symmetric plasma mode follows a square root
and the antisymmetric mode clearly shows the inverse square root dependence.
For $BSCCO$, the coupled regime is shorter than in $YBCO$, 
ranging between $0.1$ to $5 \mu m^{-1}$. 
For the maximum anisotropy of $10^3$, the coupled regime  disappears thus 
only surviving the optical and the asymptotic regimes.
We conclude that an extremely  large anisotropy inhibits
coupling between the two surfaces and enhances the surface  plasma modes. 

Fig.(\ref{fig3}), directly obtained from Eq.(\ref{omas}), 
shows the asymptotic frequency versus
$\tilde \varepsilon$, for several anisotropies.
Notice that the asymptotic frequency for small $\tilde \varepsilon$
is just $\omega_p$.
For $YBCO$, the asymptotic frequency drops over an order of magnitude
when $\tilde \varepsilon$ changes by three decades.
The choice of a non-conducting medium of small dielectric constant can
render impossible the observation of these plasma modes because the
asymptotic frequency becomes larger than the superconducting gap.
In $BSCCO$, and other extremely anisotropy compounds, 
the asymptotic regime is always much smaller than the gap.
As $\tilde \varepsilon$ increases, 
we have determined the crossover value, $\varepsilon_s \lambda_{\perp}/\lambda_{\parallel}$, where the
assymptotic frequency acquires a $\tilde \varepsilon$ dependence, thus  
being strongly affected by  anisotropy. 
Hence in order that  the   inverse square-root dependence   
of the antisymmetric coupled regime  be observable,
the condition $\tilde \varepsilon > \varepsilon_s
\lambda_{\perp}/\lambda_{\parallel}$ must be satisfied.

Fig.(\ref{fig4}) shows the dispersion relations above $\omega_p$
for a $100nm$-thick film.
The dielectric constant $\tilde \varepsilon$ is taken equal to 
$\varepsilon_s$.
We have chosen a $BSCCO$ compound with an anisotropy
$\lambda_{\perp}/\lambda_{\parallel}$ = $10^2$.
Therefore the c-axis plasma frequency is much smaller than the
superconducting gap\cite{BEASLEY}.
The curves were directly obtained from Eq.(\ref{gapl}) and Eq.(\ref{xapl}).
Above $\omega_p$, plasma modes branch  into a large number of 
modes when film thickness increases.
For the choice of a $100nm$-thick film, 
the present London-Maxwell  theory remains valid for the first 
66 modes.
We  plot in this figure the  $M=1,2,3,4,7,12$ and $66$ dispersion relations. 
For increasing $M$ the plasma mode  becomes slower,
indicating that the number of reflections that the confined plane wave 
undergoes at  the interfaces
has also  increased. 
Notice that the $M=66$ plasma mode is very near $\omega_p$.
The M=1 branch is very near to the optical branch  showing 
that it essentially has  the speed of light in
the dielectric medium.

\section{Conclusion}\label{section5}

In this paper  we have studied propagating plasma modes in
a superconducting film surrounded by two 
identical dielectric media. 
The superconductor is anisotropic having its uniaxial
direction (c-axis) perpendicular to the interfaces
with the dielectric medium.
We consider the existence of a plasma frequency along the uniaxial direction
below the gap and study its effects into the propagating
modes using the London-Maxwell theory,
which gives a good account of the
physical situation for scales larger than the interplane
separation. 
In fact we do use the interplane separation as a cut-off
for the present theory.
We only consider low incident frequency compare to the superconducting gap
frequency. Moreover the wavelength associated to the incident wave is
much larger than London's penetration length
along the surfaces.
Therefore the dielectric constant of the superconducting film along the plane
is always negative and with a large  modulus.
Along the c-axis, the situation is quite distinct.
The dielectric constant of the superconducting film along this direction
vanishes at the plasma frequency,
being positive above.
In this paper we were only concerned about
modes that propagate parallel to the surfaces and evanesce in the
dielectric medium.
Under this condition, and the requirement of the non-conducting medium
dielectric constant larger than the superconductor's static constant, 
we find that modes are either
totally above the plasma frequency, or below,
thus never crossing the plasma frequency line.
Thus we study these two regions of frequency separately.
Within our London-Maxwell theory framework we
find the {\it exact} expressions for the dispersion
relations of the plasma modes in the two cases.
The exact expressions are found in parametric form,
the parameter being the one that characterizes the 
exponential behavior inside the superconductor.
Approximated expressions for some especial regimes
of the dispersion relations are obtained from
our exact parametric solution.
In this way we retrieve well-known results in the
literature\cite{MOOIJ,MIRHASHEN,MISHONOV2,DPB,ARTEMENKO2} to the 
context of an anisotropic
film with a plasma frequency inside, and also,
 derive new ones.
We find that propagating modes below and above the
plasma frequency have quite distinct physical
properties.

Below the plasma frequency, the amplitude also evanesce
inside the superconductor, the film thickness is
important in order to assure a sufficiently strong
coupling between the surfaces.
There are two branches of the dispersion relation
corresponding to the two possible arrangements
of the superficial charge densities.
The symmetric branch is the lowest in frequency
and has the two superficial charge density symmetrically
disposed.
The highest branch has opposite charge facing each
other  at the interfaces thus being
antisymmetric.
We have studied here in details the 
three possible regimes that can exist for these two branches,
their cross-over, and the conditions for the existence of
the coupled regime, the most interesting of the three regimes.
In the so-called  coupled regime both symmetric and
antisymmetric   dispersion 
relation branch provides independent information on the
two London penetration lengths.
The antisymmetric mode is intimately connected to the transverse current 
component,
thus being highly sensitive to the transverse London penetration
length,  in the same fashion that  the 
symmetric mode  depends on the
longitudinal London penetration length.
We believe that this  remarkable property can be used to gain
further understanding of the transverse current component
in anisotropic and layered films.

Above the plasma frequency the amplitudes do not
evanesce inside the film, propagation is geometrically
understood, and an oblique ray can represent
the plane wave inside the superconductor which is totally
reflected at
the interfaces (see Fig.(\ref{fig1})).
However outside the film, in the dielectric medium,
the wave remains evanescent, and we conclude that
above the plasma frequency the 
superconducting film displays confined propagation,
similar to an optic fiber.
Notice that the film thickness is not a crucial parameter
because no matter how apart the surfaces are, there will
always be plane waves travelling in its interior.
Contrary to below the plasma frequency, above there are
many modes corresponding essentially to the number of
half-wavelengths that fit inside the film along the
c-axis.

In  summary we have shown in this paper that 
superconducting films surrounded by a 
dielectric medium displays very interesting 
plasma modes propagation properties   at  frequencies  
below the gap frequency and such properties are quite distinct
below and above the c-axis plasma frequency.

This work was done under a CNRS(France)-CNPq(Brasil) collabotation
program.

\newpage

\section{Appendix}\label{apen} 

In this appendix we provide further
details on how the exact parametrized solutions of section \ref{section3}
leads the the pictures developed in sections 
\ref{section2} and section \ref{section3}
for below and above the plasma frequency, respectively.

\vskip0.5truecm
\par \noindent $\underline{\omega < \omega_p}$ \quad

The three possible regimes, optical coupled and assymptotic 
follow from the exact parametrized solution.
In particular we show that Eq.(\ref{eqs}) and Eq.(\ref{eqa}), 
for  the coupled regime, and Eq.(\ref{omas}), 
for the asymptotic regime, follow from the Eq.(\ref{gbpl}) and  
Eq.(\ref{xbpl}).
We derive both Eq.(\ref{eqs}) and Eq.(\ref{eqa}) within
the following approximations:
{\it (i)} retardation effects are neglected ($\gamma \ll 1$)
($\sqrt{1-\gamma^2}\approx 1$);
{\it (ii)} fields evanesce slowly inside the film, 
$t \ll 1$. Therefore the Eq.(\ref{tab3}) dispersion relations become
$\sqrt{A\;X} \gamma^2 \approx  1/2$  and 
$\sqrt{A\;X} \gamma^2 \approx 2/t^2$  for the symmetric and
for the antisymmetric modes, respectively; and
{\it (iii)} the film thickness is much smaller than the London
penetration length along the surfaces $A\gg 1$ so that
even though $t$ is small we can approximate Eq.(\ref{eqt}) by
$A\, t^2 \approx \sqrt{ X\,R/(1-\gamma^2\,X\,R/r )}$.
Direct elimination of the parameter $t$ leads to 
both Eq.(\ref{eqs}) and Eq.(\ref{eqa}).
In the asymptotic regime fields inside the superconductor fade away 
very quickly from the surfaces , 
thus corresponding to  $t \gg 1$.    
The  surface decoupling is seen from the antisymmetric and  symmetric 
dispersion relations which converge to the same 
asymptotic frequency $\omega_{as}$. 
According to  Eq.(\ref{gbpl})
$g_{A}(t\rightarrow \infty ) = 
g_{S}(t\rightarrow \infty )\rightarrow t^2$ 
and  Eq.(\ref{omas}) follows in a straighforward way from this
argument.

\vskip0.5truecm
\par \noindent $\underline{\omega > \omega_p}$ \quad
 
To derive Eq.(\ref{obap}) from the exact parametrized solution
is in fact very simple, because it does not demand a detailed
study of the $\gamma(t')$ curve.
To do so we reparametrize the exact solution of  Eq.(\ref{gapl}),
replacing $t'$ by a new parameter
$t_1$,  $t' = t_{I} \,- t_1$ where $t_S= (2\,N\,+1)\pi$, and 
$t_A = 2\,(N+1)\,\pi$. The advantages are two-fold:
both symmetric and antisymmetric modes are now defined in the
same interval $0 \le t_1 \le \pi$; and
the approximation is under control, 
corresponding to the limit  $ t_1/t_{I} \ll 1$.
Notice that at $t_1=0$ all curves satisfy the condition $\gamma =1$,
thus justifying our claim that Eq.(\ref{obap}) is a
good approximation for the exact parametrized dispersion relation of
Eq.(\ref{gapl}) and Eq.(\ref{xapl}) when the phase velocity
is approximately given by the speed of light in the dielectric.
Indeed to obtain Eq.(\ref{obap}) from Eq.(\ref{gapl}) and Eq.(\ref{xapl}), 
notice that $\gamma(t_{I}-t_1)$ can be Taylor expanded
in powers of  $t_1$ because the functions $h_{I}(t')$, introduced in section
\ref{section3}, are well behaved in this neighborhood: 
$h_{I}(t') = (t_{I} \,- t_1)^2 \tan^2{(t_1/2)}$.
We introduce the approximation $ t_1/t_{I} \ll 1$ into Eq.(\ref{xapl})
obtaining that
$X_{I} \approx (r/R)/\{\gamma_{I}^2 - r/[A(t_{I})^2+1)]\}$.
The key issue here  is that all the parametric 
dependence of $X$ is now on $\gamma$, because the last term
was approximated by a constant, $r/A(t_{I}-t_1)^2 \approx
r/A t_{I}^2$.
Therefore we  obtain the approximated parametrized 
dispersion relation,
\begin{eqnarray}
\omega(t_1) &\approx & \omega_p \; \sqrt{{{ \gamma_{I}(t_1)^2  }\over
{ \gamma_{I}(t_1)^2-{{r}\over{A\,t_{I}^2}} }} } \\
\beta(t_1) &\approx & \beta_c \;
\sqrt{{1\over{ \gamma_{I}(t_1)^2-{{r}\over{A\,t_{I}^2}}}}}
 \end{eqnarray}
from where the family of curves $\omega(\beta)$ of Eq.(\ref{obap}) 
follow, by  suitably removing  the function $\gamma(t1)$.

Finally we would like to get some more detailed information
about the parametrized exact solution, in particular the two 
following issues:{\it (i)} $\omega(\beta)$ 
is always an increasing function of $\beta$;
{\it (ii)} near $\omega_p$ it suffices to  consider the positive root
of Eq.(\ref{gapl}) because the positive and negative roots of  
Eq.(\ref{gapl}) are just parts of the same curve and meet  where
the square root vanishes.
The claims are  of general validity but we derive them
under some further working assumptions on Eq.(\ref{gapl}),
that in leading order, acquires a very simple form,
\begin{eqnarray}
A\,(t_{I}-t_1)^2 \gg r \gg 1, \quad 
\gamma_{I}(t_1)^2 = {{1\pm\sqrt{1-4 {{r^2}\over{R}}\, \tan^2{({{t_1}\over {2}})}}}
\over{ 2\lbrack 1 + 
{{A\,r}\over{R}}\,(t_{I}-t_1)^2 \,\tan^2{({{t_1}\over {2}})}   \rbrack}}
\label{gapl2}
\end{eqnarray}
To have the condition $r\gg 1$ satisfied one must choose a 
dielectric medium of  sufficiently high constant:
$\tilde\varepsilon \gg \varepsilon_s$.
To have  at least one mode described by  the above inequality,
one must require that the minimal frequency satisfies
the inequality: $t'_{max} \gg \sqrt{r/A}$, which is
$\sqrt{ \tilde \varepsilon/\varepsilon_s} \ll 2\pi
\lambda_{\parallel}/a $.
To show that the curve never saturates,
take  $t_1=0$, where one finds that
$\gamma=1$ and $\gamma=0$ for the positive and negative roots, 
respectively. 
Obviously the positive root is the low frequency part of
the curve.
Notice that   the negative root is not physical
because for $\gamma=0$ $\beta(X)$ is negative according to
Eq.(\ref{xapl}). 
This just means that the negative square root curve 
must end  at a finite non-zero $\gamma$,
where the denominator of
Eq.(\ref{xapl}) vanishes and so $X$  diverges.
$\gamma$ is limited between zero and one, so it must
also be finite at this point.
Consequently $\omega$  must also diverge, and so, we conclude from 
this that the curve never saturates in $\omega$ or $\beta$.

\newpage

    
\newpage 

\baselineskip = 2\baselineskip  
\begin{figure}
\caption{ A pictorial view of wave propagation in a superconducting film 
surrounded by two equivalent non-conducting media. For 
$\omega < \omega_p$ the instantaneous electric field and the
superficial charges are shown for symmetric (a) and antisymmetric (b) modes. 
For $\omega > \omega_p$ the optical ray associated to  the plane wave
travelling inside the film is shown here (c) for M=1,2,3 and 4. 
The symmetry of each state is also shown here,  
for both cases (a), (b) and (c),
through the diagram $E_z$ versus $x$. }
\label{fig1}
\end{figure}
\begin{figure}
\caption{ The symmetric and antisymmetric $\omega < \omega_p$ 
dispersion relations are shown for three anisotropies.
A very thin superconducting film, $d=10nm$-thick, surrounded by $SrTiO_3$, 
is considered.}
\label{fig2}
\end{figure}
\begin{figure}
\caption{ This Figure shows the asymptotic frequency versus
$\tilde \varepsilon$, for three anisotropies.
For small $\tilde \varepsilon$  the asymptotic frequency
is just $\omega_p$.
The asymptotic frequency for $YBCO$ is the most sensitive
to changes in the susbstrate dielectric constant.}
\label{fig3}
\end{figure}
\begin{figure}
\caption{The dispersion relations above $\omega_p$ is shown
for a $100nm$-thick film and anisotropy
$\lambda_{\perp}/\lambda_{\parallel}$ = $100$.
The $M=1$ mode propagates with speed $v$ and at the $M=66$ mode is
the upper limit for the validity of the present theory.}
\label{fig4}
\end{figure}

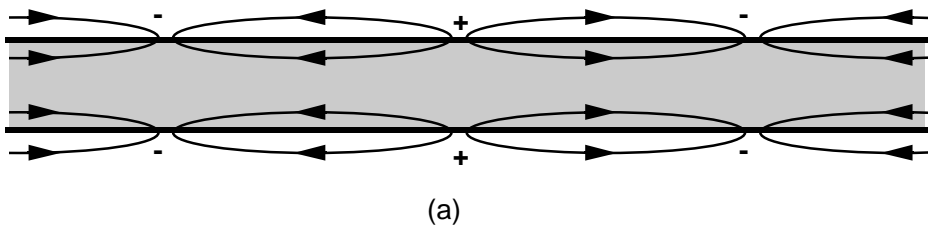
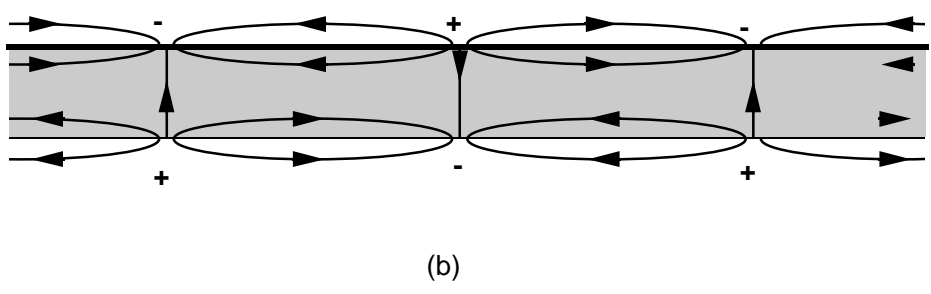
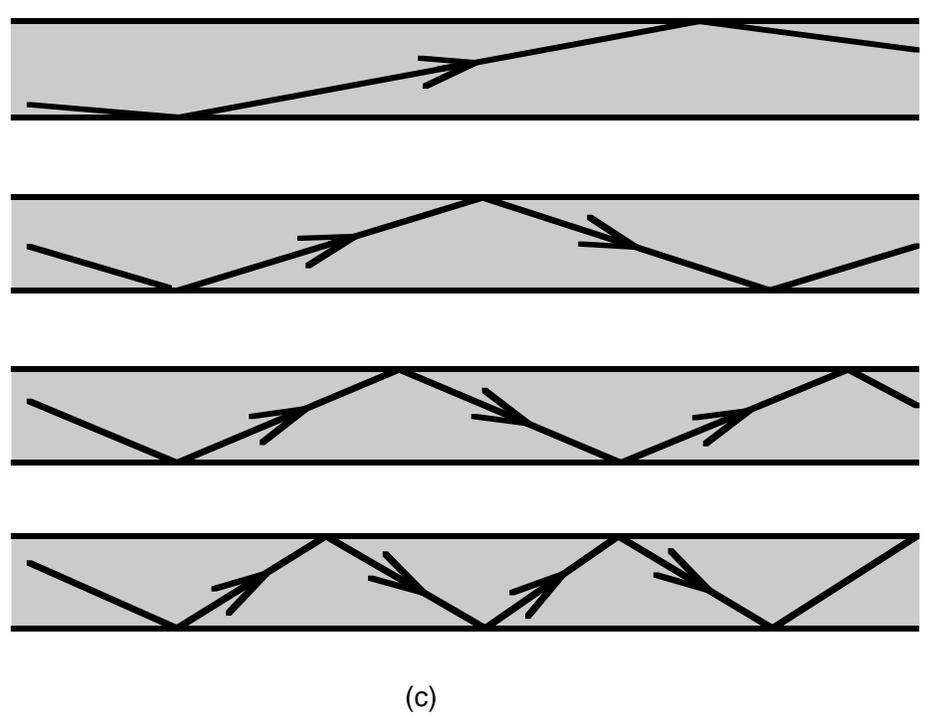
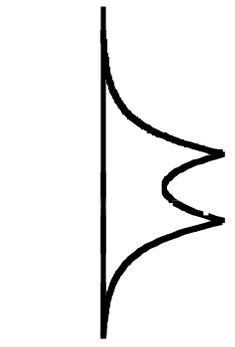
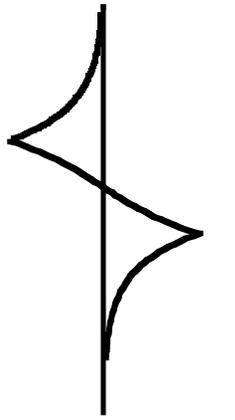
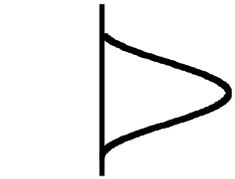
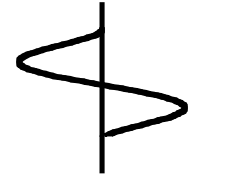
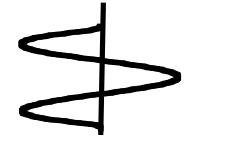
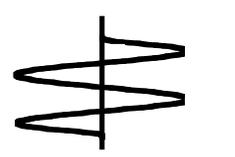
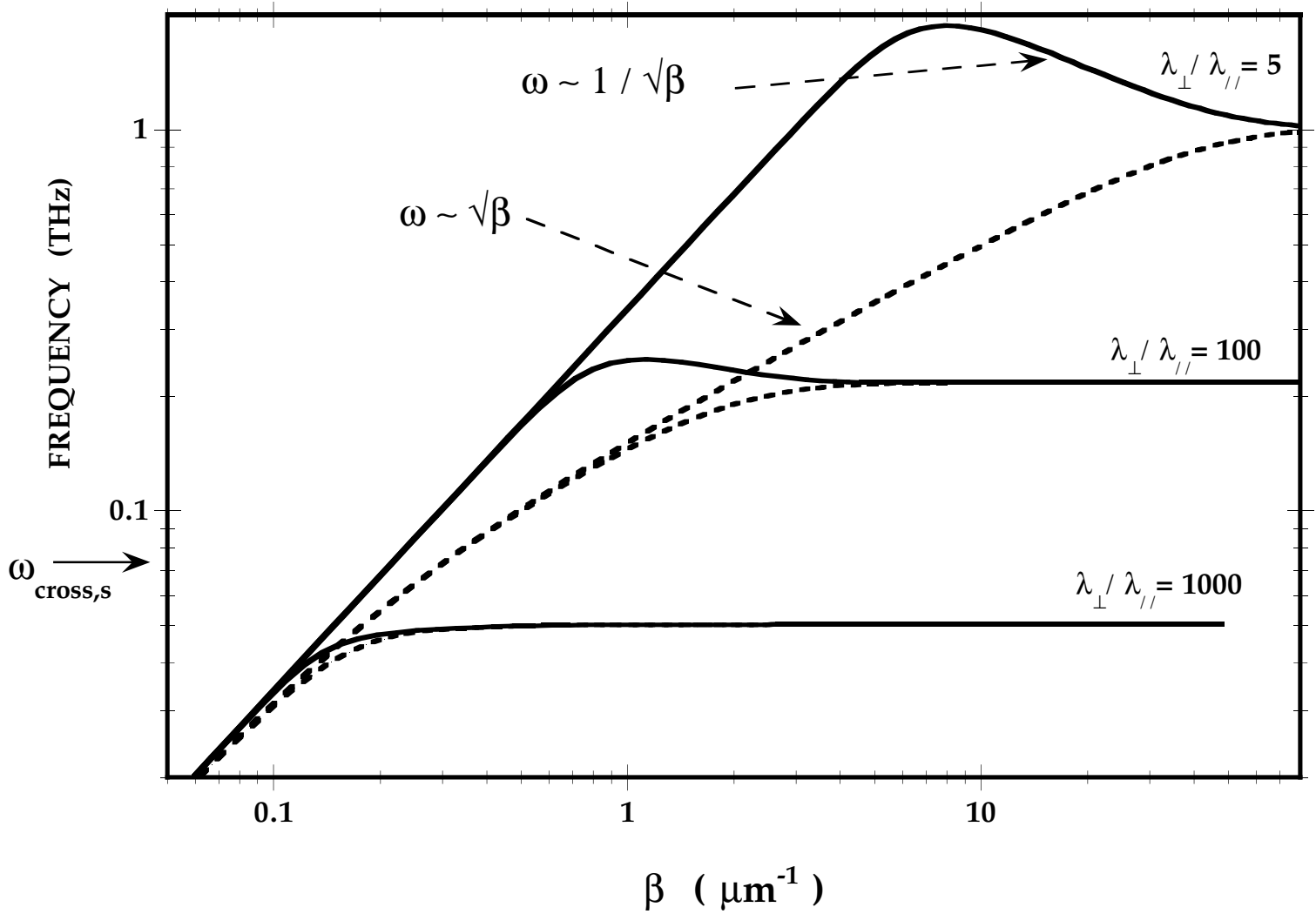
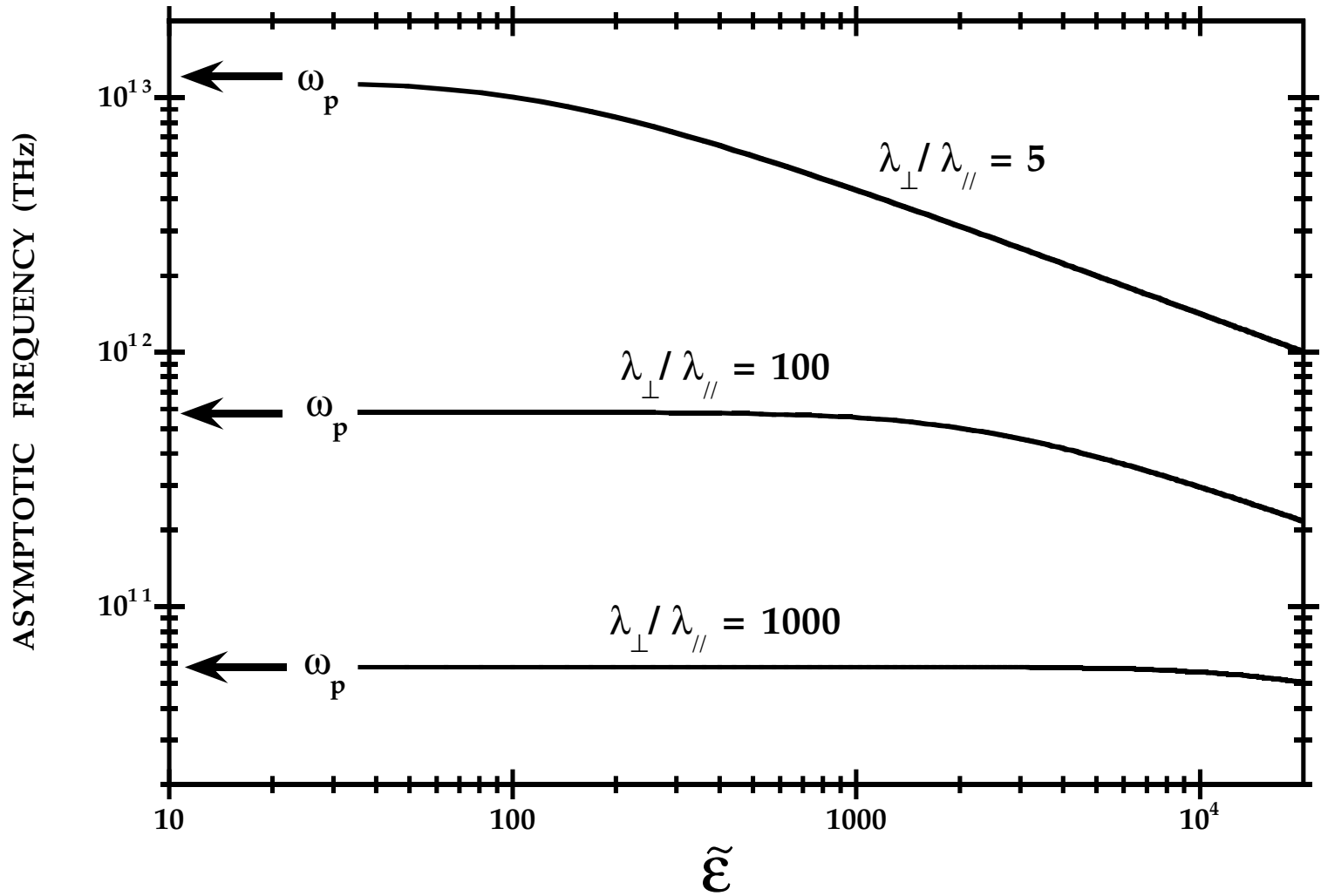
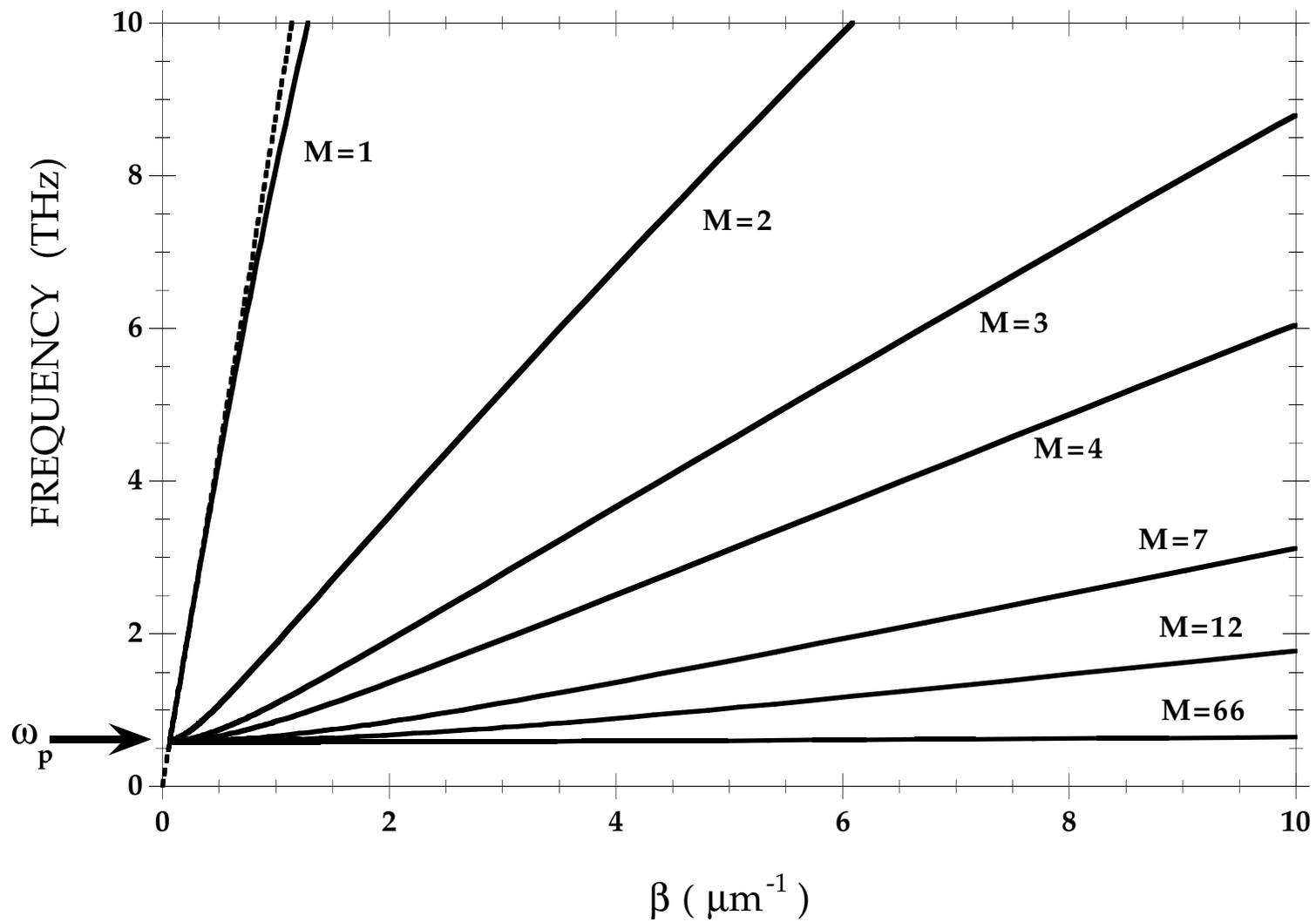
\end{document}